\documentclass{revtex4}                    
\usepackage{amsmath}
\usepackage{times}
\usepackage{amssymb}
\usepackage{graphicx}
\usepackage{float}
\usepackage{color}
\usepackage[utf8]{inputenc}
\newcommand{\la}{\left\langle}
\newcommand{\ra}{\right\rangle}
\newcommand{\rme}{{\rm e}}
\newcommand{\rmd}{{\rm d}}
\newcommand{\rmi}{{\rm i}}

\begin{document}
\today
\title{Anomalous long-range correlations at a non-equilibrium phase transition}
\author{Antoine Gerschenfeld and Bernard Derrida}
\affiliation{ Laboratoire de Physique Statistique, Ecole Normale
Sup\'erieure, UPMC Paris 6,Universit\'e Paris Diderot, CNRS,
24 rue Lhomond, 75231 Paris Cedex 05 - France}
\begin{abstract}

Non-equilibrium diffusive systems are known to exhibit long-range correlations, which decay
like the inverse $1/L$ of the system size $L$ in one dimension. Here, taking the example of
the $ABC$ model, we show that this size dependence becomes anomalous (the decay becomes a
non-integer power of $L$) when the diffusive system approaches a second-order phase
transition.

This power-law decay as well as the $L$-dependence of the time-time correlations can be
understood in terms of the dynamics of the amplitude of the first Fourier mode of the particle
densities. This amplitude evolves according to a Langevin equation in a quartic potential,

 which was introduced in a previous work to explain the anomalous behavior of the cumulants of
the current near this second-order phase transition. Here we also compute some of the
cumulants away from the transition and show that they become singular as the transition is
approached.

\end{abstract}

\maketitle

\section*{Introduction}
Long range correlations are a well known property of non-equilibrium systems in their steady
state\cite{Spohn,SC,DKS,GLMS,OS,DELO,BDGJL-cor}. For one dimensional diffusive systems, which
satisfy Fourier's law, these correlations (between pairs of points separated by a macroscopic
distance) scale like the inverse $1/L$ of the system size $L$\cite{Spohn,BDGJL-cor,DLS5}. They
can also be related to the non local nature of the the large deviation functional of the
density profiles\cite{BDGJL-cor,derrida2007}.

These long range correlations have been calculated in a few cases, using in
particular fluctuating hydrodynamics\cite{Spohn,GLMS,DELO,BDLV}. 
It was observed that when a diffusive system approaches a second order phase transition, the
factor in front  of these $1/L$ correlations becomes singular\cite{BDLV}.
The goal of the present work is to analyse these long range correlations
at and in the neighborhood of a phase transition. To do so, we consider the $ABC$ model,
one of the simplest diffusive systems which undergoes a phase transition.

The main result obtained in the present work is that the long range correlations at the
second-order phase transition of the $ABC$ model have an anomalous dependence on the system
size which can be understood by an effective theory \cite{GD} for the amplitude of the slow density mode
which becomes unstable at the transition.

This effective theory, which was developed in a recent work \cite{GD}, led us to predict an
anomalous size dependence of the cumulants of the particle current at the transition. Here,
we also show that, away from the transition, these cumulants satisfy Fourier's law (i.e. are
proportional to $1/L$), with prefactors which become singular at the transition.

The outline of the paper is as follows: in section \ref{sec:abc}, we first recall some known
properties of the $ABC$ model as well as our previous results\cite{GD} on the cumulants of the
current, which exhibit an anomalous Fourier's law at the second-order transition. In sections
\ref{sec:relax} and \ref{sec:steady}, we show how the effective theory for the slow density
mode yields a $L^{-1/2}$ size dependence of the correlation function in the critical regime.
In section \ref{sec:cur}, we calculate the cumulants of the particle current (all the
cumulants in the flat phase and the first two cumulants in the modulated phase). They all 
decay like $1/L$, with prefactors which become singular at the transition, and which match
the expressions obtained directly in the critical regime in \cite{GD}.

\section{Short review of the ABC model}\label{sec:abc}

The $ABC$ model is a one-dimensional lattice gas, where each site is occupied by one of three
types of particles, $A$, $B$ or $C$. Neighboring sites exchange particles at the rates
\begin{align}
	AB & \underset{1}{\overset{q}{\rightleftharpoons}} BA\nonumber\\
	BC & \underset{1}{\overset{q}{\rightleftharpoons}} CB\label{rates}\\
	CA & \underset{1}{\overset{q}{\rightleftharpoons}} AC\nonumber
\end{align}
with an asymmetry $q\leq 1$. This model has been studied on a closed and on an open interval
(with particle reservoirs at each end)\cite{ACLMMS,LM,LCM,BLS} as well as on a ring with
periodic boundary conditions\cite{EKKM1,EKKM2,CDE,FF1,FF2,CM,BCP,BD,GMS}: in this article, we
consider the latter case by studying a ring of $L$ sites.

For $q=1$, all configurations are equally likely in the steady state; on the other hand, for
$q<1$, the particles of the same species tend to gather \cite{EKKM1}. When $q$ scales as
$$q=e^{-\beta/L}\,,$$
the dynamics of the model becomes diffusive\cite{HS,BDGJL3} : for each species $a\in\{A,B,C\}$, one can define
a rescaled density profile $\rho_a$,
\begin{equation}\label{profils}
    {\rm Pro}[\mbox{site } k \mbox{ is of type }a] = \rho_a(k/L,t/L^2) \equiv \rho_a(x,\tau)\,. 
\end{equation}
Because the microscopic dynamics \eqref{rates} conserves the numbers of particles, the
$\rho_a(x,\tau)$ are related to their associated currents $j_a(x,\tau)$ by the conservation
laws
\begin{equation}\label{jrho}
    \partial_t\rho_a(x,\tau) = -\partial_x j_a(x,\tau)\,.
\end{equation}
By assuming that the underlying microscopic regions of the system are local equilibrium
\cite{CDE,BDLV,BD2005}, one can show that the currents $j_a(x,\tau)$ satisfy noisy, biased Fick's laws:
\begin{equation}\label{jfluc}
    j_a = -\partial_x\rho_a + \beta\rho_a(\rho_c-\rho_b) + {1\over\sqrt{L}}\eta_a(x,\tau)
\end{equation} 
where $b$ and $c$ designate the previous and next species with respect to $a$, and where
the $\eta_a(x,\tau)$ are Gaussian white noises such that
\begin{multline*}\label{sigma}
    \la \eta_a(x,\tau)\eta_{a'}(x',\tau')\ra = \sigma_{aa'}\delta(x-x')\delta(\tau-\tau')
    \\\mbox{ with } \sigma_{aa'} = \left\{ \begin{array}{ll}
        2\rho_a(1-\rho_a) &\mbox{ if } a=a'\\
        -2\rho_a\rho_{a'} &\mbox{ otherwise}
    \end{array}\right.
\end{multline*}
(note that these correlations imply $\eta_A+\eta_B+\eta_C=0$ due to the fact that $j_A
+j_B+j_C$ is identically zero). At leading order in $L$, the noise in \eqref{jfluc} can be
neglected, so that \eqref{jrho} leads to the evolution equations\cite{CDE}
\begin{equation}\label{hydrodet}
    \partial_\tau \rho_a = \partial_x^2\rho_a + \beta\partial_x\rho_a(\rho_b-\rho_c)\,.
\end{equation}
For low $\beta$, the flat density profiles $\rho_a(x,\tau) = r_a$ are a stable solution of
these equations. They become linearly unstable\cite{EKKM1,CDE,CM} when $\beta$
reaches $\beta_*$ given by
$$\beta_* = {2\pi\over\sqrt{\Delta}} \;\;\mbox{ with }\;\; \Delta=1-2\sum_a r_a^2\,`,$$
indicating a second-order transition at $\beta = \beta_*$ : above $\beta_*$, the steady-state
profiles are modulated. It has been argued that these steady-state density profiles are
time-independent\cite{CM,BD}.

This stability analysis does not rule out the possibility of a first-order phase transition
taking place at some $\beta_t < \beta_*$. A more detailed analysis of the neighborhood of
$\beta_*$ shows that for $\Lambda<0$, the transition should be first order\cite{CDE}, with
\begin{equation}\label{Lambda}
	\Lambda = \sum r_a^2 - 2\sum r_a^3\,.
\end{equation}
While no analytical expression for $\beta_t$ is known in this case, it has been studied
numerically in \cite{CM}.

In \cite{GD}, we considered the $ABC$ model in the region $\Lambda>0$ where the
transition is expected to be second order. We found that, for a system of size $L$, there
exists a critical regime $|\beta-\beta_*|\sim 1/\sqrt{L}$ for which the dynamics of the
density profiles is dominated by those of their first Fourier mode.
Let
\begin{equation}\label{defRa}
    R_A(t) = {1\over L}\sum_{k=1}^L \rme^{-2\rmi\pi k/L} A_k(t) \mbox{ where }
    A_k(t) = \left\lbrace \begin{aligned}
        1 &\mbox{ if site } k \mbox{ is of type } A\\
        0 &\mbox{ otherwise}
    \end{aligned}\right.
\end{equation}
be the first Fourier mode of the density of species $A$. Then, near the transition, the
first Fourier modes of the other densities, $R_B(t)$ and $R_C(t)$, are related to $R_A$ by
\begin{equation}\label{Ri}
	\left\{ \begin{array}{l}
		R_B(t) = {2r_C-1-\rmi\sqrt{\Delta}\over2r_A}R_A(t)\\
		R_C(t) = {2r_B-1+\rmi\sqrt{\Delta}\over2r_A}R_A(t)
	\end{array}\right.\,,
\end{equation}
and the evolution of $R_A(t)$ can be described by a Langevin equation in a quartic potential
in terms of the diffusive time $\tau = t/L^2$:
\begin{equation}\label{evolfluc}
    {\rmd R_A\over \rmd\tau} = 4\pi^2\left[\gamma - {2\Lambda\over r_A\Delta^2} |R_A|^2\right]R_A
    +{\mu_A(\tau)\over\sqrt{L}}
\end{equation}
with
\begin{equation}\nonumber
    \gamma = {\beta-\beta_*\over \beta_*}
\end{equation}
and with $\mu_A$ is a complex Gaussian white noise:
$$\la \mu_A(\bar\tau)\mu_A^*(\bar\tau')\ra = 24\pi^2 {r_A^2r_B r_C\over\Delta}\delta(\tau-\tau')\,.$$ 
By rescaling $R_A(\tau)$ by
\begin{equation}\label{deff}
    R_A(\tau) = \left[\Delta r_A^3 r_B r_C\over \Lambda L\right]^{1/4} f(\bar\tau)\,,
\end{equation}
one can see that \eqref{evolfluc} becomes
\begin{align}\label{rescale}
	 {\rmd f\over \rmd\bar\tau} = (\bar\gamma-|f(\bar\tau)|^2)f(\bar\tau) + \mu(\bar\tau)\,,
\end{align}
where  $\la \mu(\bar\tau)
\mu^*(\bar\tau')\ra = \delta(\bar\tau-\bar\tau') $ and with
\begin{align}\label{deftg}
	\bar\tau=8\pi^2{\sqrt{3\Lambda r_A r_B r_C}\over \Delta^{3/2}}{t\over L^{5/2}}
	\mbox{ and }
	\bar\gamma = \sqrt{L}{\Delta^{3/2}\over 2\sqrt{3\Lambda r_A r_B r_C}}{\beta-\beta_*
	\over \beta_*}\,.
\end{align}
Hence, in the critical regime $|\beta-\beta_*| \simeq 1/\sqrt{L}$, the amplitude $R_A$ of the
first Fourier mode varies on a time scale $\bar\tau \propto t/L^{5/2}$, with an amplitude in
$1/L^{1/4}$.

In \cite{GD}, we showed that, due to these slow fluctuations of $R_A(t)$, the integrated
particle current $Q_A(t)$ of $A$ particles during time $t$ through a section in the system
exhibits anomalous fluctuations at the transitions, reminiscent of those that can be
numerically observed in momentum-conserving mechanical models\cite{BDG}:
\begin{align}\label{qscale}
    \la Q_A(t) \ra &\simeq {t\over L} \beta r_A(r_C-r_B) + {t\over L^{3/2}} A_1\\
    \la Q_A^n(t) \ra &\simeq {t\over L^{5/2-n}} A_n \mbox{ for } n \geq 2\,\nonumber.
\end{align}
We also argued that the coefficients $A_n$ can be expressed in terms of
$n$-point correlation functions of the solution $f(\bar\tau)$ of \eqref{rescale}:
\begin{equation}\label{defcn}
    C_n(\bar\gamma) = \lim_{\bar\tau\to\infty}{1\over\bar\tau}\int_0^{\bar\tau}d\bar\tau_1..
    d\bar\tau_n \la |f(\bar\tau_1)..f(\bar\tau_n) |^2\ra_c\,.
\end{equation}
Therefore in \cite{GD} we reduced the calculations of the cumulants of $Q_A(t)$ in the 
critical regime to the study of the Langevin equation of a single particle evolving in a
quartic potential.

\section{Power-law relaxation of the first Fourier mode}\label{sec:relax}
\begin{figure}[b]
	\centerline{\includegraphics{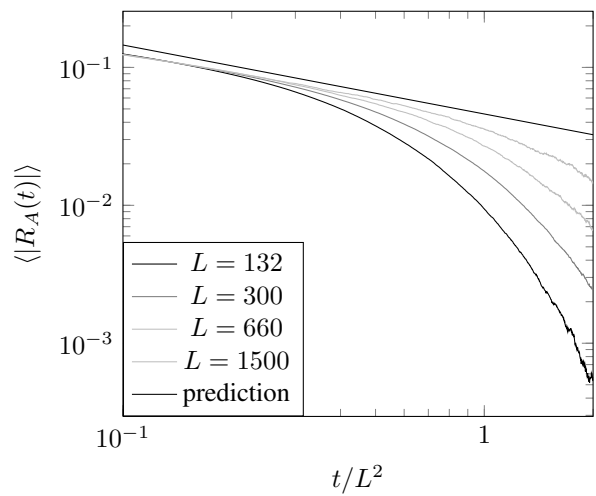}}
    \caption{\label{fig:decay}Relaxation of the first-mode amplitude $|R_A(t)|$ measured using
    \protect\eqref{defRa} for systems of size $132\leq L \leq 1500$ started in a segregated
    initial configuration, for $r_A=r_B=r_C=1/3$ and $\beta=\beta_*$. Our analysis
    \protect\eqref{Radecay} of the first mode of the densities predicts a $t^{-1/2}$ decay
    (see \protect\eqref{Radecay}).}
\end{figure}

In this section, we study the decay of the first Fourier mode in the critical regime when its
initial amplitude is much larger than its steady-state values $(\sim L^{-1/4})$. This is the
case when one starts from a non steady-state initial condition: here we consider such a
relaxation, starting from a fully segregated initial configuration of the type
$AA..AABB..BBCC..CC$.

According to \cite{GD}, we expect all the higher Fourier modes of the densities to relax on
the hydrodynamic time scale $\tau = t/L^2$. After this initial relaxation, the density
profiles \eqref{profils} should be dominated by a first Fourier mode with amplitude $R_A(t)$
evolving according to \eqref{evolfluc},\eqref{deff}; moreover, as long as $R_A$ remains much
larger than $L^{-1/4}$, the noise term in \eqref{evolfluc} can be neglected and the evolution
of $R_A$ reduces to
\begin{equation}\label{evoldet}
    {\rmd R_A\over \rmd\tau} = 4\pi^2\left[\gamma - {2\Lambda\over r_A\Delta^2}|R_A|^2\right] R_A\,.
\end{equation}
When $\beta =\beta_*$ (i.e. when $\gamma=0$), $R_A$ should thus decay as a power law:
\begin{equation}\label{Radecay}
	R_A(\tau) = {R_A(0)\over \sqrt{1+{16\pi^2\Lambda\over r_A \Delta^2}|R_A(0)|^2 \tau}}
	\underset{\tau\to\infty}{\simeq} {R_A(0)\over |R_A(0)|}{\Delta\over 4\pi}\sqrt{r_A\over \Lambda \tau}\,.
\end{equation}
In Figure \ref{fig:decay}, we measured numerically the amplitude $|R_A|$ from its definition
\eqref{defRa} for systems of $132\leq L\leq 1500$ particles for $r_A=r_B=r_C=1/3$ and
$\beta=\beta_*$. The power-law decay \eqref{Radecay} should be valid in a rather limited range of time
($L^2\ll t\ll L^{5/2}$), and is rather difficult to observe. Fortunately,
the higher Fourier modes seem to relax fast enough for the power-law decay to occur
already for $t/L^2\simeq 10^{-1}$: our data for increasing sizes seems to converge to the
power-law \eqref{Radecay} in the whole range $10^{-1}<t/L^2<2$.

Figure \ref{fig:decay} also shows a departure from this power law at a rescaled time
$t/L^2$ increasing with $L$ : in the next section, we show that this is due to the
increasing effect of the noise term in \eqref{evolfluc} as $R_A$ decreases.

\subsection*{Power-law decay at the tricritical line}

The damping term of the evolution equation for $R_A$ \eqref{evoldet} that we obtained around
$\beta=\beta_*$ vanishes on the tricritical line $\Lambda = 0$ (see eq. \eqref{Lambda}). In
this case, it is necessary to push the analysis of \cite{GD} further in order to obtain an
effective equation for $R_A$. One can show that, when $\Lambda=0$, \eqref{evoldet} is replaced
by
\begin{equation}\label{evoldet3}
	{\rmd R_A\over \rmd\tau} = 4\pi^2\left(\gamma-{|R_A|^4\over r_A^2\Delta^2}\right)R_A\,.
\end{equation}
Along this tricritical line ($\gamma=0$, $\Lambda=0$), the $\tau^{-1/2}$ decay of $R_A$ 
in the critical regime \eqref{Radecay} should become in $\tau^{-1/4}$. 

\section{Correlations in the steady state}\label{sec:steady}

It has been shown that the $ABC$ model exhibits long-range steady-state correlations, scaling
as $1/L$, in the flat phase $\beta<\beta_*$ \cite{BDLV}. In the modulated phase $\beta>\beta_*$, these correlations should be of order $1$ due to the modulation of the
steady-state profiles.

In this section, we show, using our effective dynamics \eqref{evolfluc},\eqref{rescale} for the first mode $R_A$
\eqref{evolfluc}, that these correlations scale as $1/\sqrt{L}$ in the critical regime; 
we also show that temporal correlations decay on the slow time scale $\bar\tau\propto
t/ L^{5/2}$ at the transition. Finally, we briefly comment on the behavior on the tricritical
line $\Lambda=0$.

\subsection{Spatial correlations in the critical regime}

\begin{figure}[t]
	\centerline{\includegraphics{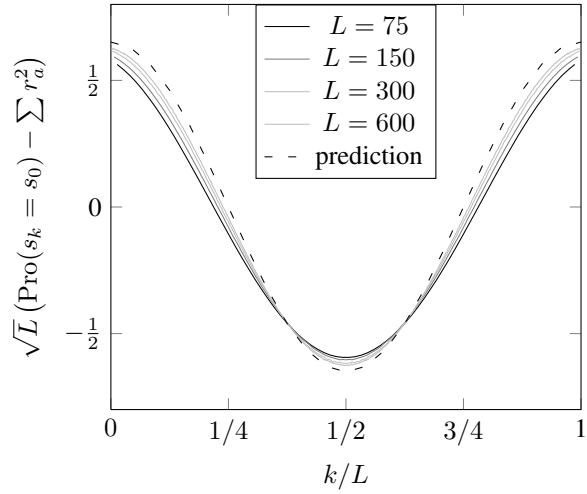}}
	\label{fig:steady}
    \caption{Steady-state density correlations ${\rm Pro}[s_k = s_0]-\sum r_a^2$ measured for systems
    of size $75\leq L \leq 600$, with $r_A=r_B=r_C=1/3$ and $\beta=\beta_*$. Our analysis of
    the first mode of the densities \protect\eqref{evolfluc} predicts cosine correlations of 
    an amplitude scaling as $1/\sqrt{L}$ \protect\eqref{steadycorr}.}
\end{figure}

As explained in \cite{GD}, the density fluctuations in the critical regime are dominated by
those of the first Fourier mode $R_A$. Thus one can calculate to leading order
the steady-state correlations of the densities at the critical point:
\begin{align}
	{\rm Pro}[s_k = s_l] - \sum_a r_a^2 &
	\simeq \sum_a \la \left(r_a + R_a \rme^{2\rmi\pi k/L}+cc.\right)\left(r_a + R_a 
	\rme^{2\rmi\pi l/L}+cc.\right)\ra _c\,.\nonumber
\end{align}
The Langevin equation \eqref{rescale} can be expressed as a
Fokker-Planck equation over the probability density of the rescaled first mode $f(\bar\tau)$,
\\$P(f_x,f_y,\bar\tau) \equiv {\rm Pro}[f(\bar\tau) \simeq f_x+if_y]$ :
\begin{equation}\label{fkp}
	{\partial P\over\partial\bar\tau} = {\rm div}\left[(r^2-\bar\gamma) P \vec r\right]
    +{1\over 4} {\nabla^2} P
\end{equation}
with $\vec r = (f_x,f_y)$ and $\bar\gamma$ as defined in \eqref{deftg}.
From this equation, it is easy to see that $f(\bar\tau)$ is isotropically distributed in the steady state:
\begin{equation}\nonumber
	P_0(\vec r,\tau) \propto\exp\left[2\bar\gamma r^2-r^4\right]\,.
\end{equation}
Therefore, $\la R_a^2\ra  = \la R_a^{*2}\ra  = 0$ and 
\begin{equation}\nonumber
	\la |R_a|^2\ra  = \sqrt{3\Delta r_A^3 r_B r_C\over \Lambda L} \la |f^2|\ra 
	\mbox{ , with } \la |f^2|\ra  = {\int_0^\infty r^3 \rme^{2\bar\gamma r^2-r^4} dr
	\over \int_0^\infty r \rme^{2\bar\gamma r^2-r^4} dr}\,.
\end{equation}
It is easy to compute
\begin{equation}\label{fc1}
	\la |f|^2\ra   = \bar\gamma + {1\over 2} {\rme^{-\bar\gamma^2}\over
	\int_{-\infty}^{\bar\gamma} \rme^{-z^2}dz}\equiv C_1(\bar\gamma)\,;
\end{equation}
this leads to
\begin{equation}\label{steadycorr}
	{\rm Pro}[s_k = s_l] - \sum r_a^2 \simeq 2 \sqrt{3\Delta r_a r_b r_c\over \Lambda L} C_1(\bar
	\gamma) \cos 2\pi{k-l\over L}\,.
\end{equation}
Therefore, steady-state correlations should scale as $1/\sqrt{L}$ in the critical regime.
In Figure \ref{fig:steady}, we compare our prediction \eqref{steadycorr} to numerical measurements of 
systems of $75\leq L\leq 600$ particles, for $\beta=\beta_*$ and $r_A=r_B=r_C=1/3$.
\ \\\ \\
{\bf Remark:} The $\bar\gamma\to-\infty$ and $\bar\gamma\to+\infty$ limits of our expression
for the equal-time correlations \eqref{steadycorr} can both be checked from known results. 

In \cite{BDLV}, exact expressions for the equal-time density correlations have been calculated
in the flat phase (\cite{BDLV}, eq. (23)): they diverge as $\beta\to\beta_*^-$, leading to
$${\rm Pro}[s_k = s_l] - \sum r_a^2  \simeq {6r_A r_B r_C \over \Delta L}{\beta_*\over
\beta_*-\beta}\cos 2\pi{k-l\over L}\,,$$
which is compatible with $C_1(\bar\gamma) \simeq -1/2\bar\gamma$ for $\bar\gamma\to-\infty$ in
\eqref{fc1}.

On the other hand, Equation (39) of \cite{CDE} establishes that the modulated steady-state
profiles in the disordered phase are, for $\beta\to\beta_*^+$, of the form
\begin{equation}\nonumber
    \bar\rho_a(x) = r_a+\Delta\sqrt{{r_a\over 2\Lambda} {\beta-\beta_*\over\beta_*}} 
    \rme^{2\rmi\pi (x-\varphi)}+cc.\,,
\end{equation}
with $x=k/L$ and
$\varphi$ an arbitrary phase. This leads to
$${\rm Pro}[s_k = s_l] - \sum r_a^2  \simeq {\Delta^2\over \Lambda}{\beta-\beta_*\over
\beta_*}\cos 2\pi{k-l\over L}\,,$$
which is compatible with $C_1(\bar\gamma) \simeq \bar\gamma$ for $\bar\gamma\to\infty$ in
\eqref{fc1} and \eqref{steadycorr}.

\subsection{Decay of the steady-state temporal correlations}

\begin{figure}[t]
	\centerline{\includegraphics{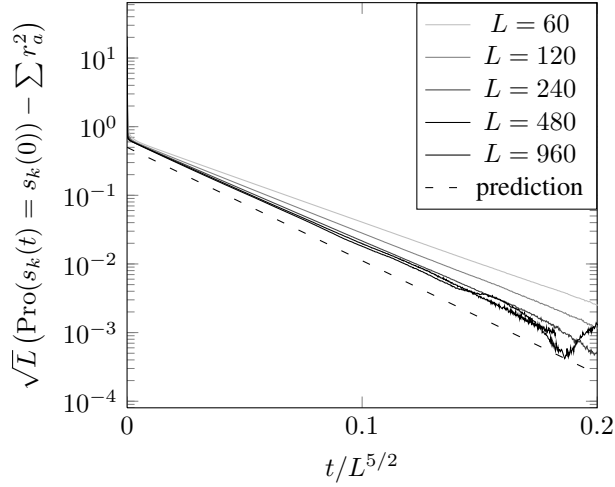}}
	\label{fig:decayexp}
    \caption{Steady-state time correlations ${\rm Pro}[s_k(t) = s_k(0)]-\sum r_a^2$ for
    systems of size $30\leq L \leq 960$, with $r_A=r_B=r_C=1/3$ and $\beta=\beta_*$, as
    functions of $t/L^{5/2}$. Our effective, fluctuating evolution equation for the first
    Fourier mode \protect\eqref{evolfluc} leads us to predict an exponential decay 
    \protect\eqref{decayexp}.}
\end{figure}

In the previous section, we have seen that, on a time scale $L^2\ll t\ll L^{5/2}$, the first
Fourier mode of the densities \eqref{defRa} should decay as a power law when the system
relaxes from a non steady state initial configuration. Here we consider the decay of the
steady-state density correlations $\la \rho_a(x,0)\rho_a(x,\tau)\ra _c$: the noisy evolution
equation \eqref{evolfluc} implies that they should decay on the "slow" time scale $t\propto
L^{5/2}$. They can be expressed in terms of the rescaled first mode $f(\bar\tau)$ as
\begin{equation}\nonumber
	{\rm Pro}[s_k(t)=s_k(0)] -\sum r_a^2\propto {1\over \sqrt{L}}\la 
	f(0)f^*(\bar\tau) + cc.\ra _c
\end{equation}
with $\bar\tau\propto \tau/\sqrt{L}$ the rescaled time \eqref{deftg}. Let $H_{\bar\gamma}$ be
the operator of the Fokker-Planck equation \eqref{fkp} over the probability density of
$f(\bar\tau)$, $P(f_x,f_y,\bar\tau)$:
\begin{equation}\nonumber
	{\partial P\over\partial\bar\tau} \equiv H_{\bar\gamma}P\,.
\end{equation}
For large $\bar\tau$, correlations such as $\la f(0)f^*(\bar\tau)\ra _c$ should decay as
$e^{\lambda_{\bar\gamma} \bar\tau}$, where $\lambda_{\bar\gamma}$ is the second largest
eigenvalue of $H_{\bar\gamma}$ (the largest being $0$). Thus, we expect time correlations to
decay on the time scale $t\propto L^{5/2}$ as
\begin{equation}\label{decayexp}
	{\rm Pro}[s_k(0) = s_k(t)]-\sum r_a^2 \propto {1\over \sqrt{L}} \exp\left[
	-\alpha {t\over L^{5/2}} \right]
\end{equation}
with $\alpha=-8\pi^2{\sqrt{3\Lambda r_A r_B r_C}\over \Delta^{3/2}}\lambda_{\bar\gamma}$.
While there is no known analytical expression for $\lambda_{\bar\gamma}$, one can determine it
numerically by approximating the operator $H_{\bar\gamma}$ over a finite subspace of
$L^2({\mathbb R}^2)$ of growing dimension. Because the steady-state (and thus the
$0$-eigenvector of $H_{\bar\gamma}$) is $P_0(f_x,f_y) \propto e^{2\bar\gamma \vec r^2-\vec
r^4}$, we consider the subspace
\begin{equation}\nonumber
    P_Q(f_x,f_y) = Q(f_x,f_y) \rme^{2\bar\gamma r^2-r^4}\,,
\end{equation}
where $Q(f_x,f_y)$ is a polynomial of degree less than some finite $N$ in $(f_x,f_y)$. We need
to define a scalar product $\la \cdot|\cdot\ra$ for which $H_{\bar\gamma}$ is Hermitian: to do
so, we choose 
\begin{align}\label{scalprod}
    \la P_Q | P_R \ra &= \iint dx dy \rme^{r^4-2\bar\gamma r^2} P_Q(x,y) P_R(x,y)\\
    &= \iint dx dy \rme^{2\bar\gamma  r^2- r^4} Q(x,y) R(x,y)\,,\nonumber
\end{align}
for which the matrix elements of $H_{\bar\gamma}$ read
\begin{equation}\label{hmatrix}
    \la P_Q | H_{\bar\gamma} | P_R \ra = -{1\over 4} \iint dx dy \rme^{2\bar\gamma 
    r^2-r^4}(\vec\nabla Q).(\vec\nabla R)\,.
\end{equation}
For $\bar\gamma = 0$, we constructed an orthonormal basis of the subspace of the $P_Q$ with
respect to the scalar product \eqref{scalprod} for $2\leq N\leq 8$; then, we computed the
matrix elements of $H_0$ \eqref{hmatrix} in this basis. The second largest eigenvalue of these
successive hermitian matrices appears to converge to
$$\lambda_0 \simeq -0.83\,.$$ 
In Figure \ref{fig:decayexp}, we compare this prediction with the results of simulations in
the steady-state for $60\leq L\leq 960$, $r_A=r_B=r_C=1/3$ and $\beta=\beta_*$.

The exponential decay \eqref{decayexp} also seems to predict the cut-off of the power law
decay of Figure \ref{fig:decay} observed in systems relaxing from an arbitrary initial
condition to the stationary state.
\subsection{Correlations on the tricritical line}

As in the deterministic case the cubic term of the fluctuating evolution equation for the
first Fourier mode \eqref{evolfluc} vanishes on the tricritical line $\Lambda = 0$. The
fluctuating correction of the deterministic tri-critical evolution equation \eqref{evoldet3}
is the same as those of \eqref{evolfluc} :
\begin{equation}\nonumber
    {\rmd R_A\over \rmd\tau} = 4\pi^2\left(\gamma-{|R_A|^4\over r_A^2\Delta}\right)R_A
    +{\mu_A\over \sqrt{L}}\,.
\end{equation}
This leads to the rescaling
$$ R_a= {\sqrt[3]{6 r_a^4r_b r_c}\over L^{1/6}} g(\tilde\tau) \mbox{ with }
{\rmd g\over \rmd\tilde\tau } = (\tilde\gamma-|g|^4) g + \mu(\tilde\tau)
\,,$$
with
$$\tilde\tau = {4\pi^2\over \Delta} (6r_ar_br_c)^{2/3}{t\over L^{8/3}} \mbox{ and } 
\tilde\gamma = {\Delta\over (6r_ar_br_c)^{2/3}} L^{2/3}{\beta-\beta_*\over \beta_*}\,.$$
Hence, the tricritical regime occurs for $|\beta-\beta_*|\propto L^{-2/3}$ : it is
characterized by fluctuations of the first Fourier mode of amplitude $L^{-1/6}$ on a time
scale $t\propto L^{8/3}$.

In its steady state, we expect spatial correlations to scale as $${\rm
Pro}[s_k(t)=s_l(t)]-\sum r_a^2 \propto {1\over L^{1/3}}\cos 2\pi{k-l\over L}\,,$$
while time correlations should decay exponentially on the time scale $t\propto L^{8/3}$, as
$${\rm Pro}[s_k(0)=s_k(t)]-\sum r_a^2 \propto {1\over L^{1/3}}\exp\left[-\alpha
{t\over L^{8/3}}\right]\,.$$

\section{Fluctuations of the current in the $ABC$ model}\label{sec:cur}

In this section, we study the cumulants of the integrated current of particles of type $A$,
$Q_A(t)$, as $\beta\to\beta_*$. 

In contrast to \cite{GD}, where we studied the critical regime $|\beta-\beta_*|\sim 
1/\sqrt{L}$, here we compute the cumulants to leading order in $L$ for a fixed $\beta\neq\beta_*$: the expressions we obtain diverge at $\beta_*$. For
$\beta\to\beta_*^-$ , we find \eqref{cdivdis}
\begin{equation}
    \left\{ \begin{array}{ll}\label{cdivdisform}
        {\la Q_A(t)\ra\over t} &\simeq {A_1\over L} + {B_1 \over L^2(\beta-\beta_*)}\\
        {\la Q_A^2(t)\ra_c\over t} &\simeq {A_2 \over L} + {B_2  \over L^2(\beta-\beta_*)^3}
    \end{array}\right.
      \mbox{ and  } {\la Q_A^n(t)\ra _c\over t} \simeq {B_n \over L^2 
      (\beta-\beta_*)^{2n-1}}
\end{equation}
while we find that, for $\beta \to\beta_*^+$ \eqref{cdivmod},
$${\la Q_A^2(t)\ra_c\over t} \simeq {B'_2\over L(\beta-\beta_*)}\,.$$
We also show that these expressions are compatible with the $\bar\gamma\to\pm\infty$ limits of
the expressions of the cumulants \eqref{c1crit},\eqref{cncrit} in terms of the functions
$C_n(\bar\gamma)$ \eqref{defcn} derived in the critical regime in \cite{GD}.

\subsection{First cumulant of the current}

We start with the assumption that the steady state density profiles $\bar\rho_a(x)$, which are
the long-time limits of the leading-order, deterministic hydrodynamic equations
\eqref{hydrodet}, are time-independent, both in the flat and in the modulated phase. The
associated particle currents $j_a$ are thus homogeneous, $j_a(x,\tau)=J_a$, and can be
computed by integrating \eqref{jfluc} in space, yielding
\begin{equation}\label{defja}
    J_a = \int_0^1 \beta \bar\rho_a(x)(\bar\rho_c(x)-\bar\rho_b(x))dx\,.
\end{equation}
Then, the average integrated current $\la Q_A(t)\ra $ through any position in the system will
behave in the long-time limit like $\la Q_A(t)\ra  \simeq {t\over L} J_A$, so that
\begin{equation}\nonumber
    {\la Q_A(t)\ra \over t} \simeq {1\over L} \int_0^1 \beta 
    \bar\rho_A(x)(\bar\rho_C(x)-\bar\rho_B(x))dx \,.
\end{equation}

For $\beta < \beta_*$, we thus obtain ${\la Q_A(t)\ra} \simeq {t\over L}\beta r_A(
r_C-r_B)$. For $\beta>\beta_*$, the analytic expression of the $\bar\rho_a(x)$ is rather
complicated\cite{CM}. However, their limit as $\beta\to\beta_*^+$ takes a simple form (see
\cite{CDE}, or the $\bar\gamma\to\infty$ limit of \eqref{evolfluc}):
\begin{equation}\label{profmod}
    \bar\rho_a(x) \underset{\beta\downarrow\beta_*}{\simeq} r_a + 
    (R_a \rme^{2\rmi \pi x} + cc.)
\end{equation}
with $|R_A| = \Delta\sqrt{r_A(\beta-\beta_*)/2\Lambda \beta_*}$ and with $R_B$, $R_C$ 
related to $R_A$ by \eqref{Ri}. This leads to an analytical expression for 
$\la Q_A(t)\ra $ when $\beta\to\beta_*^+$ :
\begin{equation}\nonumber
    {\la Q_A(t)\ra \over t} \underset{\beta\downarrow\beta_*}{\simeq} {r_C-r_B\over L}\left[
    \beta r_A - {\Delta^2\over\Lambda}(\beta-\beta_*)\right]\,.
\end{equation}

\subsection{Second cumulant of the current}

In order to compute fluctuations of the current $Q_A(t)$, the noise terms of the biaised
Fick's law \eqref{jfluc} have to be taken into account. These fluctuating hydrodynamics can be
reformulated as a large deviation principle known as the macroscopic fluctuation theory (MFT)\cite{BDGJL1,BDGJL2,BDGJL5,BD,BDLV}, which gives the probability of observing a given evolution $\rho_a(x,\tau)$ of the density
profiles :
\begin{equation}\label{mft}
    {\rm Pro}[\rho_a(x,\tau)] \propto \exp\left[-L\iint dxd\tau
    {1\over 2} (j-q).\sigma^{-1}(j-q)\right]
\end{equation}
where $j = (j_A,j_B)$, $q = (q_A, q_B)$ (with $q_a = -\partial_x\rho_a + \beta\rho_a(
\rho_c-\rho_b)$ the deterministic part of \eqref{jfluc}) and with
\begin{equation}\nonumber
    \sigma = \left(\begin{array}{rr}
        \sigma_{AA} & \sigma_{AB}\\
        \sigma_{BA} & \sigma_{BB}\\        
    \end{array}
    \right) =2\left(\begin{array}{rr}
        \rho_A(1-\rho_A) & -\rho_A \rho_B\\
         -\rho_A \rho_B &\rho_B(1-\rho_B)\\
    \end{array}\right)\,.
\end{equation}
The generating function $\log\la e^{\lambda Q_A(t)}\ra $ can then be calculated as the solution of
a variational problem :
\begin{equation}\label{qgenopt}
    \log\la e^{\lambda Q_A(t)}\ra  = \max_{\rho_a,j_a} L \int_0^{t/L^2} d\tau \int_0^1 dx
    \left[\lambda j_A - {1\over 2} (j-q).\sigma^{-1}(j-q)\right]\,.
\end{equation}
For $\lambda = 0$, the optimum in the above is the solution to the deterministic equations
\eqref{hydrodet}, which satisfy by definition $j = q$ : these are $\rho_a(x,\tau) = r_a$
(for $\beta \leq \beta_*$) or $\rho_a(x,\tau) = \bar\rho_a(x)$ (for $\beta > \beta_*$), and
$j_a(x,\tau) = J_a$ defined by \eqref{defja}.

For $\beta < \beta_*$, the optimal profiles are still flat even for $\lambda \neq 0$ (as 
we will see below, they are a stable minimum):
\begin{equation}\label{profoptdis}
    \left\lbrace \begin{aligned}
        \rho(x,\tau) & = r\\
        j(x,\tau) &= q + \sigma\left(\begin{array}{r}\lambda\\0
        \end{array}\right) \equiv J(\lambda)
    \end{aligned}\right.
\end{equation}
with $\rho=(\rho_A,\rho_B)$, $r=(r_A,r_B)$, and $q$ and $\sigma$ taking their constant values
for $\rho=r$. This leads to
\begin{equation}\label{qgendis}
    \log \la e^{\lambda Q_A(t)}\ra  \underset{\beta<\beta_*}\simeq {t\over L} \left[\beta\lambda r_A(r_C-r_B) + \lambda^2 r_A(1-r_A)\right] \equiv {t\over L}F_{\rm flat}(\lambda)
\end{equation}
and to
\begin{equation}\label{cumdis1}
    {\la Q_A^2(t)\ra _c\over t} \simeq {2\over L}r_A(1-r_A) \mbox{ and } {\la Q_A^n(t)\ra _c\over t}
    = o\left(1\over L\right) \mbox{ for } n \geq 3\,.  
\end{equation}

On the other hand, the optimal profiles in \eqref{qgenopt} vary with $\lambda$ for $\beta>
\beta_*$, with non-trivial optimization equations. We will therefore restrict ourselves to
the calculation of $\la Q_A^2(t)\ra $, for which it is sufficient to compute \eqref{qgenopt} to 
second order in $\lambda$.

In order to do so, we consider profiles close to the $\lambda=0$ optimum,
$\rho_a = \bar\rho_a(x)$ and $j_a = J_a$. Because arbitrary translations of these profiles are
also optimal, we need to consider profiles moving at a small velocity $v$ : this amounts to

$$\left\lbrace \begin{aligned}
    \rho(x,t) &= \bar\rho(x-vt)+\mu(x-vt)\\
    j(x,t) &= J + K - v\bar\rho(x-vt)
\end{aligned}\right. $$
with $\rho=(\rho_A,\rho_B)$ and $\mu(x),v,K \ll 1$. Because $q_a = -\partial_x\rho_a + 
\beta\rho_a(\rho_c-\rho_b)$, this leads to the following expression for $q$ :
$$q = \bar q  - \mu' - \bar M \mu \mbox{ with } \bar M = \beta\left(\begin{array}{rr}
    1-2\bar\rho_C & 2\bar\rho_A\\
    -2\bar\rho_B & 2\bar\rho_C-1
\end{array}\right)\,,$$ 
$\bar q = q(\bar \rho)$ and $\mu' = \partial_x\mu$. Then, the right-hand side of \eqref{qgenopt} becomes
\begin{multline}\nonumber
    S = \max_{K,v,\mu(x)} {t\over L}\int dx \left[
    \lambda(\bar J_A + K_A -v\bar\rho_A)\right. \\\left.- {1\over 2}(K-v\bar\rho+\mu'+\bar M 
    \mu)     .\bar\sigma^{-1}(K-v\bar\rho+\mu'+\bar M \mu)\right]
\end{multline}
with $\bar\sigma = \sigma(\bar\rho)$.
The optimization equations over $K$, $v$ and $\mu$ can be written as
\begin{equation}\label{optf}
    \left\lbrace \begin{aligned}
        \int dx F &= \left(\begin{array}{r}
            \lambda\\0
        \end{array}\right)\\
        \int dx \bar\rho.F & =\lambda r_A\\
        F' &= \bar M^{T} F
    \end{aligned}\right.
\end{equation}
with $F = \bar\sigma^{-1}(K-v\bar\rho+\mu'+\bar M \mu)$.
The right-hand side of \eqref{qgenopt} then reads
$$S = \max_{K,v,F(x)} {t\over L}\int dx \left[
    \lambda(\bar J_A + K_A -v\bar\rho_A)- {1\over 2}F.\bar\sigma F\right]$$
It can be expressed completely in terms of the optimal $F(x)$ solution of \eqref{optf} by
using
\begin{align}
    \int dx F.\bar \sigma F &= \int F.(K-v\bar\rho+\mu'+\bar M \mu)\nonumber\\
    &=K. \int dx F - v\int dx \bar\rho.F +\int dx \mu.(\bar M^TF-F')\nonumber\\
    &= \lambda(K_A-vr_A)\nonumber
\end{align}
so that the generating function is given to second order in $\lambda$ by
\begin{equation}\label{c2mod}
     \log\la e^{\lambda Q_A(t)}\ra  \simeq {t\over L} \int dx\left[\lambda \bar J_A 
     + {1\over 2} F.\bar\sigma F\right]
\end{equation}
By
determining the solution $F(x)$ of \eqref{optf} and integrating \eqref{c2mod} numerically,
$\la Q_A^2(t)\ra_c$ can be predicted to leading order in $L$ for $\beta > \beta_*$.
This prediction diverges as $\beta\to\beta_*^+$ : because $\bar\rho(x)$ is
known analytically in this limit \eqref{profmod}, we were able to calculate $F(x)$ exactly,
\begin{equation}\nonumber
    F(x) = {4\pi\lambda(r_B-r_C)\over \Delta^{3/2}(\beta-\beta_*)}\left(
    \begin{aligned}
        R_B\\-R_A
    \end{aligned}\right) \rme^{2\rmi\pi x}+cc.+ {\mathcal O}(1)\,,
\end{equation}
leading to a simple expression for $\la Q_A^2(t)\ra _c$ in this limit:
\begin{equation}\label{cdivmod}
    \la Q_A^2(t)\ra _c\underset{\beta\downarrow\beta_*}\simeq {t\over L}{12\pi r_A r_B r_C(r_B-
    r_C)^2\over \sqrt{\Delta}\Lambda(\beta-\beta_*)}\,.
\end{equation}

\subsection{Higher-order corrections in the flat phase}

As shown above, the macroscopic fluctuation theory \eqref{qgenopt} predicts, for large $L$,
Gaussian fluctuations of $Q_A(t)$ in the flat phase $\beta<\beta_*$, with a non-singular
variance \eqref{cumdis1} as $\beta\to\beta_*^-$.

Here, we calculate the next $1/L$ corrections to the generating function \eqref{qgendis}, by generalizing to the $ABC$ model the approach followed in \cite{ADLW} in
the case of a single conserved quantity. These corrections, of order $1/L^2$,
can be obtained by considering the large-deviation principle \eqref{mft} as a functional
integral :
\begin{equation}\label{qgenint}
    \hspace{-3mm}\la e^{\lambda Q_A(t)}\ra  \propto \int D[\rho_a]D[j_a] \exp\left[L
    \iint dxd\tau\left(\lambda j_A - {1\over 2}(j-q).\sigma^{-1}(j-q)\right)\right]
\end{equation}
where the integral takes place over all profiles $(\rho_a,j_a)$ compatible with the
conservation law $\partial_x j_a + \partial_\tau\rho_a=0$. Fluctuations around the
optimum profile \eqref{profoptdis} then give corrections to the saddle-point expression 
\eqref{qgendis}. We now consider such fluctuations, expressing them in terms of their Fourier
modes :
\begin{equation}\nonumber
    \left\lbrace \begin{aligned}
        \rho(x,\tau) &= r + \delta\rho(x,\tau)\\
        j(x,\tau) &= J(\lambda) + \delta j(x,\tau)
    \end{aligned}
    \right. \mbox{ with } \left\lbrace \begin{aligned}
        \delta\rho(x,\tau) &= \sum_{k,\omega} k\left[\alpha_{k\omega}\rme^{\rmi(kx-\omega t)} 
        + cc.\right]\\
        \delta j(x,\tau) &= \sum_{k,\omega} \omega\left[\alpha_{k\omega}\rme^{\rmi(kx-\omega 
        t)} + cc.\right]
    \end{aligned}\right.
\end{equation}
with $\alpha_{k\omega}=(\alpha_{k\omega}^{(A)},\alpha_{k\omega}^{(B)})$ the amplitude of the
fluctuations of wave number $k$ and pulsation $\omega$, which take discrete values :
\begin{equation}\nonumber
    k = {2\pi n} \mbox{ with } n\in {\mathbb N}^* \;\mbox{ and }\; \omega = {2\pi m\over 
    t/L^2} \mbox{ with } m\in {\mathbb Z}\,.
\end{equation}
Expanding \eqref{qgenint} to second order in the $\alpha_{k\omega}$, we obtain
\begin{equation}\nonumber
        \la e^{\lambda Q_A(t)}\ra  \propto \int \left[\prod_{k,\omega}d\alpha_{k\omega}
        d\alpha_{k\omega}^*\right] \exp\left[{t\over L}F_{\rm flat} (\lambda)
        -\sum_{k,\omega} \alpha_{k\omega}^* . M_{k\omega}\alpha_{k\omega}\right]
\end{equation}
with $F_{\rm flat}(\lambda)$ the dominant-order generating function \eqref{qgendis} and\\
$M_{k\omega} = {1\over 2}\left(\begin{array}{cc}
    \phi_A+\phi_C & \phi_C - 6\rmi\beta k^3\\
    \phi_C + 6\rmi\beta k^3 & \phi_B+\phi_C\end{array}\right)$, with
\begin{equation}\nonumber
\phi_a = {(J_a(\lambda)k-\omega r_a)^2\over r_a^3} + {k^4\over r_a} +\beta^2 k^2(4-9 r_a)\,.
\end{equation}
Integrating the Gaussian variables $\alpha_{k\omega}$ gives the corrections
\begin{equation}\nonumber
    \log \la e^{\lambda Q_A(t)}\ra  \simeq {t\over L} F_{\rm flat}(\lambda)
    -\sum_{k,\omega} \log P_k(\lambda,\omega) + C
\end{equation}
with $P_k(\lambda,\omega) = \phi_A\phi_B + \phi_B\phi_C+\phi_C\phi_A-36\beta^2k^6$ and $C$ an
additive constant fixed by the condition $\log\la e^{\lambda Q_t}\ra =0$ for $\lambda=0$. In 
the $t\to\infty$ limit, the sum over $\omega$ can be replaced by an
integral :
\begin{equation}\nonumber
    \log \la e^{\lambda Q_A(t)}\ra  \simeq {t\over L} F_{\rm flat}(\lambda)
    -{t\over 2\pi L^2}\sum_k\int d\omega \log P_k(\lambda,\omega) + C\,.
\end{equation}
Because only the first mode of the fluctuations, $k=2\pi$, becomes unstable as $\beta\to
\beta_*^-$, we expect the divergence in the cumulants at the transition to only affect
this $k=2\pi$ term. Thus we take the following limit in $P_{2\pi}(\lambda,\omega)$ 
: $\lambda \simeq 0$ (to determine the cumulants), $\omega\simeq 0$ (we expect slow
fluctuations to be responsible for the divergence), and $\beta\simeq\beta_*$. Then
$P_{2\pi}(\lambda,\omega)$
takes the simplified expression
\begin{equation}\nonumber
    P_{2\pi}(\lambda,\omega) \simeq 64\pi^4\left[{\omega^2+4\pi^2\Delta(\beta-\beta_*)^2
    \over r_A r_B r_C} + 12\beta_*^3(r_C-r_B)\lambda\right]\equiv Q(\lambda,\omega)\,. 
\end{equation}
The integral of $\log Q(\lambda,\omega)$ is apparently divergent : however, since
\begin{align*}
    \int_{-\Omega}^\Omega d\omega \log Q(\lambda,\omega) \underset{\Omega\to\infty}\simeq &
    4\Omega\log\Omega\\&-4\pi^2\sqrt{{\Delta (\beta-\beta_k)^2} + {12k
    \over \Delta^{3/2}}r_A r_B r_C(r_B-r_C)\lambda}\,,
\end{align*}
its divergent part is canceled out by adjusting $C$ so that $\log\la e^{\lambda Q_t}\ra =0$ 
for $\lambda=0$.
Hence, the part of the generating function which becomes singular as $\beta\to\beta_*$ reads
\begin{equation}\nonumber
    \log \la e^{\lambda Q_A(t)}\ra  \simeq {t\over L} F_{\rm flat}(\lambda)
    +{4\pi^2\gamma t\over L^2}\left[\sqrt{1 + {24\pi
    \over \Delta^{5/2}}{r_A r_B r_C\over(\beta-\beta_*)^2}(r_B-r_C)\lambda}-1\right]\,,
\end{equation}
leading to a divergence of the $n$-th cumulant of $Q_A(t)$ scaling as \eqref{cdivdisform}, 
with
\begin{equation}\label{cdivdis}
    B_n = 
    \sqrt{\Delta\over \pi }\Gamma(n-1/2)\left[
    {24\pi r_A r_B r_C(r_C-r_B)\over \Delta^{5/2}}\right]^n \,.
\end{equation}

\subsection{Anomalous fluctuations in the critical regime}

In \cite{GD}, we found that the fluctuations of the first Fourier mode on the slow time scale
$\bar\tau\propto t/L^{5/2}$ \eqref{deftg} lead to anomalous fluctuations of the integrated
current of $Q_A(t)$ in the critical regime \eqref{qscale}. More precisely, we derived
\begin{align}
        \la Q_a(t)\ra  &\simeq {t\over L}\beta r_a(r_c-r_b) + {2t\over L^{3/2}}\beta(r_b-r_c)
        \sqrt{3\Delta r_a r_b r_c\over \Lambda}C_1(\bar\gamma)\label{c1crit}\\
        \la Q_a^n(t)\ra _c &\simeq {t\over L^{5/2-n}} {8\pi^2\sqrt{3\Lambda r_a r_b r_c}
        \over \Delta^{3/2}}\left[\Delta^{3/2}(r_b-r_c)\over2\pi\Lambda\right]^n 
        C_n(\bar\gamma)\label{cncrit}
\end{align}
with $\bar\gamma$ and $C_n(\bar\gamma)$ as defined in \eqref{deftg} and \eqref{defcn}. 
Because $f(\bar\tau)$ evolves in a quartic potential \eqref{rescale}, only $C_1(\bar\gamma)$
can be easily calculated \eqref{fc1}. However, \eqref{c1crit} and \eqref{cncrit} predict
the dependence of the cumulants in $(r_A,r_B,r_C,\beta)$ in terms of the unique
parameter $\bar\gamma$ \eqref{deftg}: one can easily check that \eqref{cdivdis} and
\eqref{cdivmod} are consistent with this dependence, for
$$C_n(\bar\gamma) \underset{\bar\gamma\to-\infty}{\simeq} {(-1)^n \Gamma(n-1/2)\over
4\sqrt{\pi} \bar\gamma^{2n-1}} \mbox{ and } C_2(\bar\gamma)
\underset{\bar\gamma\to\infty}{\simeq} {2\over\bar\gamma}\,.$$

\section{Conclusion}

In this paper we have shown that the long-range correlations \eqref{steadycorr} of the $ABC$
model near the second-order phase transition decay like the $L^{-1/2}$ power of the system
size $L$. In the entire critical regime, these correlations \cite{EKKM2,BD,BDG} can be
understood from the evolution \cite{GD} of the amplitude of the first Fourier mode given by the Langevin equation of a particle in a quartic potential \eqref{rescale}.

We have also computed the cumulants of the current of particles \eqref{cdivdisform} away
from the transition, showing that the become singular at the transition in a way which
matches the results of our previous work \cite{GD} where these cumulants were computed in the
critical regime.

It would be interesting to see whether other diffusive systems, at a phase transition, display
correlation functions and current fluctuations with behaviors similar to those we discovered
here for the $ABC$ model.

Deterministic one-dimensional systems, in particular those which conserve momentum, are known
to exhibit an anomalous Fourier's law \cite{LLP2}, with cumulants of the current \cite{BDG} and correlations \cite{LMP,DLLMP,GDL} scaling as a non-integer power of the system size.
Although these systems are much more difficult to study than the $ABC$ model (for which
one only needs to follow the dynamics of a single mode), it would be interesting to see
whether the approximations that have been used so far, such as the mode-coupling approach
\cite{DLRP,LuS,hvB}, could predict the power-law dependence of these current and density fluctuations.
 
{\it Acknowledgments.}
 BD acknowledges the support of
the French Ministry of Education through the ANR 2010 BLAN 0108 01 grant.

\end{document}